\documentclass{aastex63}
\usepackage{graphicx}
\usepackage{epstopdf}
\usepackage{float}
\usepackage{color}
\usepackage{lineno}

\newcommand{\lt}{}
\newcommand{\yf}{}
\newcommand{\ltn}{}

\begin{document}





\title{Data-constrained MHD simulation for the eruption of a filament-sigmoid system in solar active region 11520}



\author{Tie Liu}
\affiliation{School of Astronomy and Space Science, Nanjing University, Nanjing, 210023, P.R. China.}
\affiliation{Key Laboratory for Modern Astronomy and Astrophysics (Nanjing
University), Ministry of Education, Nanjing, 210023, P.R. China.}

\author{Yuhong Fan}
\affiliation{High Altitude Observatory, National Center for Atmospheric Research, Boulder, CO, United States}

\author{Yingna Su}
\affiliation{Key Laboratory of Dark Matter and Space Astronomy, Purple Mountain Observatory, CAS, Nanjing 210023, P.R. China}
\affiliation{School of Astronomy and Space Science, University of Science and Technology of China, Hefei, Anhui 230026, P.R. China}

\author{Yang Guo}
\affiliation{School of Astronomy and Space Science, Nanjing University, Nanjing, 210023, P.R. China.}
\affiliation{Key Laboratory for Modern Astronomy and Astrophysics (Nanjing
University), Ministry of Education, Nanjing, 210023, P.R. China.}

\author{Ya Wang}
\affiliation{Key Laboratory of Dark Matter and Space Astronomy, Purple Mountain Observatory, CAS, Nanjing 210023, P.R. China}
\affiliation{School of Astronomy and Space Science, University of Science and Technology of China, Hefei, Anhui 230026, P.R. China}

\author{Haisheng Ji}
\affiliation{Key Laboratory of Dark Matter and Space Astronomy, Purple Mountain Observatory, CAS, Nanjing 210023, P.R. China}
\affiliation{School of Astronomy and Space Science, University of Science and Technology of China, Hefei, Anhui 230026, P.R. China}


\begin{abstract}
The separation of a filament and sigmoid is observed during an X1.4 flare on July 12, 2012 in solar active region 11520, but the corresponding magnetic field change is not clear. We construct a data-constrained magnetohydrodynamic simulation of the filament-sigmoid system with the flux rope insertion method and magnetic flux eruption code, which produces the magnetic field evolution that may explain the separation of the low-lying filament and high-lying hot channel (sigmoid). The initial state of the magnetic model contains a magnetic flux rope with a hyperbolic flux tube, a null point structure and overlying confining magnetic fields. We find that the magnetic reconnections at the null point make the right footpoint of the sigmoid move from one positive magnetic polarity (P1) to another (P3). The tether-cutting reconnection at the hyperbolic flux tube occurs and quickly cuts off the connection of the low-lying filament and high-lying sigmoid. In the end, the high-lying sigmoid erupts and grows into a coronal mass ejection, while the low-lying filament stays stable. The observed double J-shaped flare ribbons, semi-circular ribbon, and brightenings of several loops are reproduced in the simulation, where the eruption of the magnetic flux rope includes the impulsive acceleration and propagation phases.


\end{abstract}



\section{Introduction}
Solar flares and coronal mass ejections (CMEs) are the most violently energy release and explosive phenomena in the solar corona and may affect the space environment around the Earth if the releasing magnetized plasma and energetic particles reach the Earth \citep[e.g.,][]{Webb1994SoPh,Yashiro_2004_JournalofGeophysicalResearch}. It is well accepted that the explosions are driven by free energy stored in the nonpotential magnetic field of the corona. However, the coronal magnetic field is so complex that the physical mechanisms of energy release in a specific solar eruption are not yet clear. Different from the potential magnetic field which is the lowest energy state, the nonpotential magnetic field possesses free energy and can become unstable. Two kinds of nonpotential magnetic structures are generally proposed to be the pre-eruption magnetic configuration in the solar corona: sheared arcades 
and magnetic flux ropes (MFR) \citep[e.g.,][]{DeVore2000ApJ,Amari2003ApJ}, which are the three dimensional extensions of two dimensional magnetic structures proposed by \cite{Kippenhahn1957ZA} and \cite{Kuperus1974A}. The observed filaments and prominences are generally believed to be located in the dips of sheared arcades and magnetic flux ropes, which agrees with the results from magnetohydrodynamic (MHD) simulations \citep[e.g.,][]{Xia_2014,Xia_2016,Fan:2017,Fan_2018}. \cite{Ouyang2017ApJ} study 571 erupting filaments and find that 89\% are inverse-polarity filaments supported by MFRs and 11\% are normal-polarity filaments supported by sheared arcades. 


The photospheric shear and converging ﬂows \citep{vanBallegooijen_1989_apj} as well as the emergence of helical flux tubes \citep{Okamoto_2009_apj} are proposed as possible mechanisms for the formation of sheared and twisted nonpotential coronal magnetic fields. The ideal MHD instabilities and eruption mechanisms of the sheared arcade and MFR have been well studied in recent decades. \cite{Toeroek_2004_aap} show that the force-free coronal loop will be unstable (kink instability) when its twist exceeds a critical value.  \cite{Kliem_2006_PhysicalReviewLetters} have studied torus instability of a toroidal current ring and found that the MFR can erupt with a nonlinear expansion if the background field declines fast enough. In addition to MHD instabilities, magnetic reconnections are also related to eruption mechanisms. In the breakout model proposed by \cite{Antiochos_1999_apj}, magnetic reconnection at the overlying null point can remove the overlying confining field and lead to the eruption of the sheared core flux in quadrupolar magnetic fields. The tether-cutting model proposed by \cite{Moore_2001_apj} points out that in bipolar sheared fields the magnetic reconnection at the legs of the sheared arcades can increase the twist of the magnetic field and drive the eruption. \cite{Jiang2021NatAs} have confirmed the tether-cutting model in three-dimensional MHD simulations. The ideal MHD instabilities have also been examined by MHD simulations \citep[e.g.,][]{TK2005ApJ,TK2007AN,Fan2007ApJ,Aulanier2010}.

Nevertheless, the real coronal magnetic field is more complex than the ideal sheared arcade or MFR, so the physical mechanisms of the observed eruptions are also more difficult to discern. With the development of high temporal and spatial resolution equipments, more complex nonpotential magnetic structures such as the double-decker filaments \citep{Liu2012apj59a} and fan-spine magnetic fields \citep{Sun2013ApJ} are reported. However most of the aforementioned MHD simulations have been carried out with idealized magnetic field models without considering the realistic coronal magnetic fields. Recently, MHD simulations with observed data as initial and boundary conditions have sprung up and played an increasingly important role in studying real solar eruptions associated with complex magnetic structures, which are called data-constrained or data-driven MHD simulations. \cite{Jiang2013ApJ} have constructed a data-constrained MHD simulation with a complex magnetic configuration containing an MFR and fan-spine structures and reproduced some observational characteristics. With magnetic models constructed by the flux rope insertion method \citep{vanBallegooijen_2000_apj_539_983-994,vanBallegooijen_2004_apj_612_519-529}, \cite{Kliem2013} have tested stable and unstable MFRs in data-constrained simulations. Later, \cite{Inoue2014ApJ} and \cite{Amari2014Natur} have also carried out MHD simulations involving observation data and improved our understanding of solar eruptions. \cite{Guo2019} have carried out data-constrained and data-driven MHD simulations for the same solar eruption and found that the results of the two cases are very similar to each other, which suggests that the data-driven effect may be more important in  long-term buildup phase than in short-term dynamic eruptive phase.

Observations of soft X-ray (${Yohkoh}$) and full-disk H${\alpha}$ telescopes, as well as the Atmospheric Imaging Assembly (AIA) aboard the Solar Dynamics Observatory (SDO; \citealt{Lemen_2012_solphys}) show a phenomenon that low temperature filaments are located below the high temperature sigmoids. \citep[e.g.,][]{Pevtsov2002SoPh,Cheng_2016_apjs_225_16,Liu_2018}. These paired low-lying filament and high-lying sigmoid with the same chirality are called filament-sigmoid systems in this work. The pre-flare magnetic models \citep{Liu_2018} for four flare/CME events constructed by the flux rope insertion method show that the magnetic structures of the filament-sigmoid system correspond to an MFR with a hyperbolic flux tube (HFT). The filament is located below the HFT and the sigmoid corresponds to the MFR above the HFT. In this work, we carry out a data-constrained MHD simulation with a magnetic model similar to the unstable model of the first event in \cite{Liu_2018} as the initial magnetic field configuration and analyze the eruption of the filament-sigmoid system. In Section \ref{sec: description}, we introduce the observations and data-constrained MHD simulation. The results are shown in Section \ref{sec: Results}. And Section \ref{sec: Summary} presents the summary and discussion.

\section{Observations and Model Descriptions} \label{sec: description}

\subsection{Observation overview} \label{subsec: ob}
\begin{figure*}
	\centering \includegraphics[width=7in]{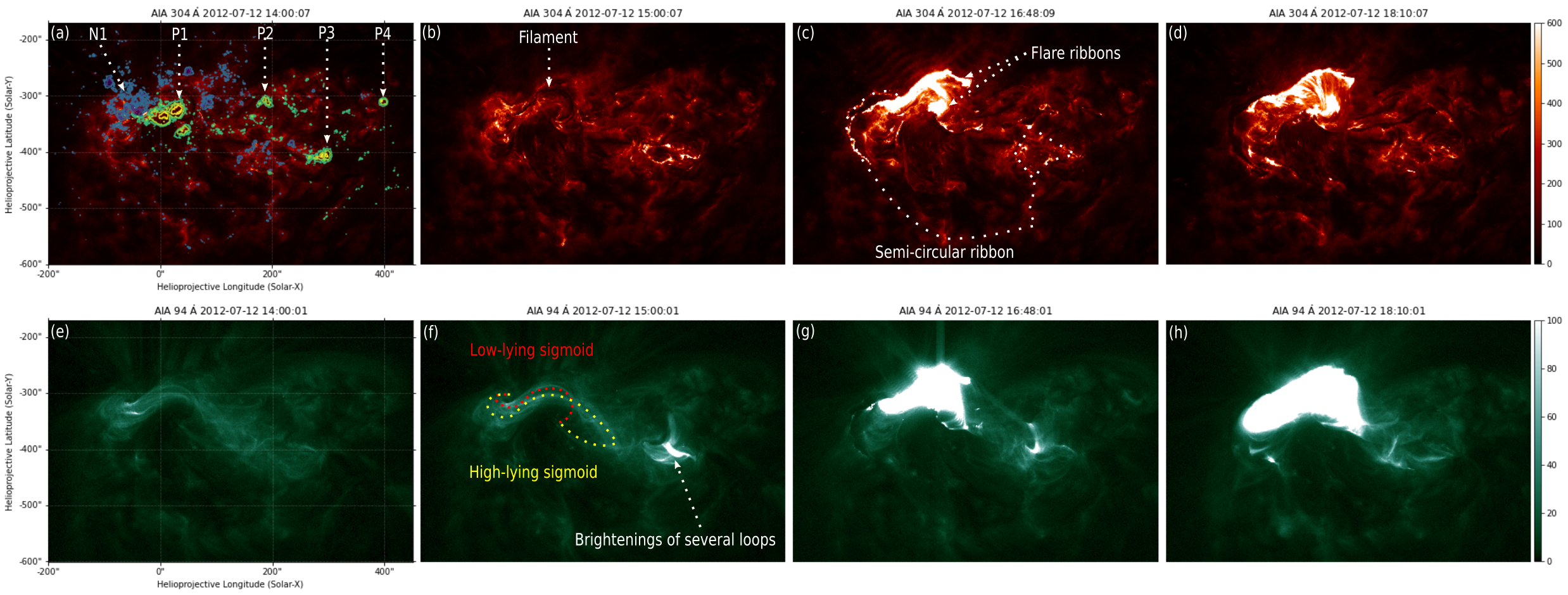}
	\caption{Multi-wavelength observations of the filament-sigmoid system. The AIA 304 Å and 94 Å images are displayed. In panel (a), the contours mark the line of sight magnetic field strength of [-1500, -1000, -500, 500, 1000, 1500] G. The main positive and negative magnetic polarities are marked by P1, P2, P3, P4 and N1. The filament is located between N1 and P1 in panel (a) and marked with a white arrow in panel (b). We trace the low-lying and high-lying sigmoids with the red and yellow dashed curves, and mark the brightenings of several loops with a white arrow in panel (f). The flare ribbons and semi-circular ribbon are marked in panel (c).} 
	\label{fig: f10} 
\end{figure*}

The observations of AIA show that an X1.4
class flare and a halo CME are produced in active regions 11520 on July 12, 2012. The soft X-ray flux of the flare starts to rise at $\sim$ 14:50 UT and peaks at $\sim$ 16:49 UT.
SDO/AIA images with a sufficiently large field of view are displayed in Figure \ref {fig: f10}, including the complex active regions 11520 and 11521. We are interested in the filament-sigmoid system (in active region 11520), which is the corresponding preflare structure of the flare and CME. Observations suggests that the eruption of the filament-sigmoid system is affected by the neighbouring active region 11521 \citep{Dudik_2014_apj_784_144}. The filament-sigmoid system contains a low-lying filament marked with a white arrow in panel (b) and a high-lying sigmoid traced with a yellow dashed curve in panel (f) similar to \cite{Cheng_2016_apjs_225_16}. In AIA 94 Å, a low-lying sigmoid marked with the red dashed curve in panel (f), lying along the filament is identified, similar to \cite{Cheng_2014_apj_789_93}. The low-lying sigmoid is also regarded as the first signature (at about 2012-07-12 15:00 UT) of the flare emission in \cite{Dudik_2014_apj_784_144}, who point out that the sigmoid develops into a highly sheared bundle of flare loops during the next 40 minutes.

The two footpoints of the filament are located in N1 and P1 in Figure \ref {fig: f10}(a), which are also the footpoints of the low-lying and high-lying sigmoids. From panels (f) and (g) of Figure \ref {fig: f10}, we find that the high-lying sigmoid extends southward and westward from P1 to P3. A semi-circular ribbon is detected in AIA 304 Å (panel (c) of Figure \ref {fig: f10}). Shortly before the eruption of the high-lying sigmoid, the brightenings of several loops appears around P3 in Figure \ref {fig: f10}(f). \cite{Dudik_2014_apj_784_144} also find that the high-lying sigmoid undergoes expansion in the southwest direction and erupts around 16:25 UT. Along the semi-circular ribbon, we detect that some small flare loops are activated in Figures \ref {fig: f10}(g) and \ref {fig: f10}(h), with the extension of the flare ribbons in Figures \ref {fig: f10}(c) and \ref {fig: f10}(d). In summary, the semi-circular ribbon, brightenings of several loops, expansion of the high-lying sigmoid, and extension of the flare ribbons are detected.

\cite{Cheng_2014_apj_789_93} consider the two sigmoids as two hot channels \citep{Zhang_2012_NatureCommunications_3_747,Cheng_2013_apj_763_43,Cheng_2014_apj_780_28} and suggests that the low-lying hot channel associated with the filament and the high-lying hot channel constitute a double-decker MFR system. 
\cite{Liu_2018} study the same active region and find that the observed filament-sigmoid system is consistent with an MFR with an HFT, hereinafter referred to as HFT-MFR system. As shown by the magnetic structure in Figure \ref {fig: f1}(e), the HFT-MFR system suggests that the high-lying sigmoid corresponds to the MFR above the HFT and the low-lying filament is located in the sheared arcades below the HFT. While in the double-decker MFR system \citep{Cheng_2014_apj_789_93}, the low-lying and high-lying sigmoids are considered as two hot channels in two MFRs above the same polarity inversion line (PIL) and the filament is located in the dips of the low-lying MFR. By contrast, here the low-lying filament is considered to be located in the magnetic dips of the sheared arcades below the HFT and the low-lying sigmoid is just the brightenings along the footpoints of the HFT due to the magnetic reconnections. \cite{Dudik_2014_apj_784_144} point out that the low-lying loop (sigmoid) lies along the outer edge of the filament, which is a signature supporting the explanation of the HFT-MFR system. In order to distinguish different magnetic field models and understand observational characteristics, coronal magnetic field construction and data-constrained MHD simulation are carried out in this study, and the corresponding results are described below.

\subsection{Model description} \label{subsec: model}

In this section, we describe the MHD simulation that models the eruptive phase of the active region evolution. The simulation uses the magnetic flux eruption code to solve the following semi-relativistic MHD equations in spherical geometry \citep{Fan:2017}:
\begin{equation}
\frac{\partial \rho}{\partial t}
= - \nabla \cdot ( \rho {\bf v}) ,
\label{eq:cont}
\end{equation}
\begin{eqnarray}
\frac{\partial ( {\rho \bf v})}{\partial t}
&=& - \nabla \cdot \left ( \rho {\bf v} {\bf v} \right )
- \nabla p + \rho {\bf g} + \frac{1}{4 \pi}
( \nabla \times {\bf B} ) \times {\bf B}
\nonumber \\
& & + \frac{{v_A}^2/c^2}{1 + {v_A}^2/c^2} \left[ {\cal I}
- {\hat{\bf b}} {\hat{\bf b}} \right ]
\cdot \left [ ( \rho {\bf v} \cdot \nabla ) {\bf v} + \nabla p - \rho {\bf g}
- \frac{1}{4 \pi} ( \nabla \times {\bf B} ) \times {\bf B} \right ] ,
\label{eq:motion}
\end{eqnarray}
\begin{equation}
\frac{\partial {\bf B}}{\partial t}
= \nabla \times ({\bf v} \times {\bf B}),
\label{eq:induc}
\end{equation}
\begin{equation}
\nabla \cdot {\bf B} = 0,
\label{eq:divb}
\end{equation}
\begin{equation}
\frac{\partial e}{\partial t} = - \nabla \cdot
\left ( {\bf v} e \right ) - p \nabla \cdot {\bf v}
- \nabla \cdot {\bf q}_s + H_{\rm num} ,
\label{eq:energy}
\end{equation}
\begin{equation}
p = \frac{\rho R T}{\mu},
\label{eq:state}
\end{equation}
\begin{equation}
e = {\frac{p}{\gamma - 1} }.
\label{eq:eint}
\end{equation}

In the aforementioned equations, ${\bf v}$ and ${\bf B}$ are the velocity field
and the magnetic field, respectively.
$\rho$, $p$, and $T$ denote respectively the plasma density, pressure and
temperature, $e$ denotes the internal energy density, $c$ is the
(reduced) speed of light, $v_A = B/{\sqrt {4 \pi \rho}}$ is the Alfv{\'e}n speed,
${\cal I}$ is the unit tensor, ${\hat {\bf  b}} = {\bf B} / B$ denotes the unit
vector in the magnetic field direction, ${\bf g} = - ( G M_{\odot} / r^2 )
{\hat {\bf r}}$
is the gravitational acceleration from the Sun. $R$, $\mu$, and $\gamma$
denote respectively the gas constant, the mean molecular weight (assuming fully
ionized hydrogen with $\mu = 0.5$), and the adiabatic index of
the perfect gas. We assume a low adiabatic index ${\gamma} = 1.1$ for
the corona plasma, which allows it to maintain its high temperature
without an explicit coronal heating.
The internal energy equation (eq. \ref{eq:energy}) takes into
account the following non-adiabatic effects:
${\bf q}_s$ refers to the field-aligned heat conduction flux, which
is computed by the hyperbolic heat conduction approach
\citep[][see equations (15) and (16) in that paper]{Rempel:2017,Fan:2017}.
In addition, we also evaluate the effective heating resulting from
the numerical diffusion of velocity and magnetic fields in the momentum and
induction equations, and add it to the internal energy in
equation (\ref{eq:energy}) as $H_{\rm num}$. Note the thermodynamics treatment in
the current simulation is different from that in \citet{Fan:2017}. Here we do not
include an empirical coronal heating nor the optically thin radiative loss, and use an
adiabatic index of $\gamma = 1.1$ instead of $\gamma = 5/3$. As a result we do not find
the formation of prominence/filament condensations that are due to the radiative instability driven
by the optically thin radiative loss as was obtained in \citet{Fan:2017}.

\begin{figure}[htb!]
	\centering \includegraphics[width=7in]{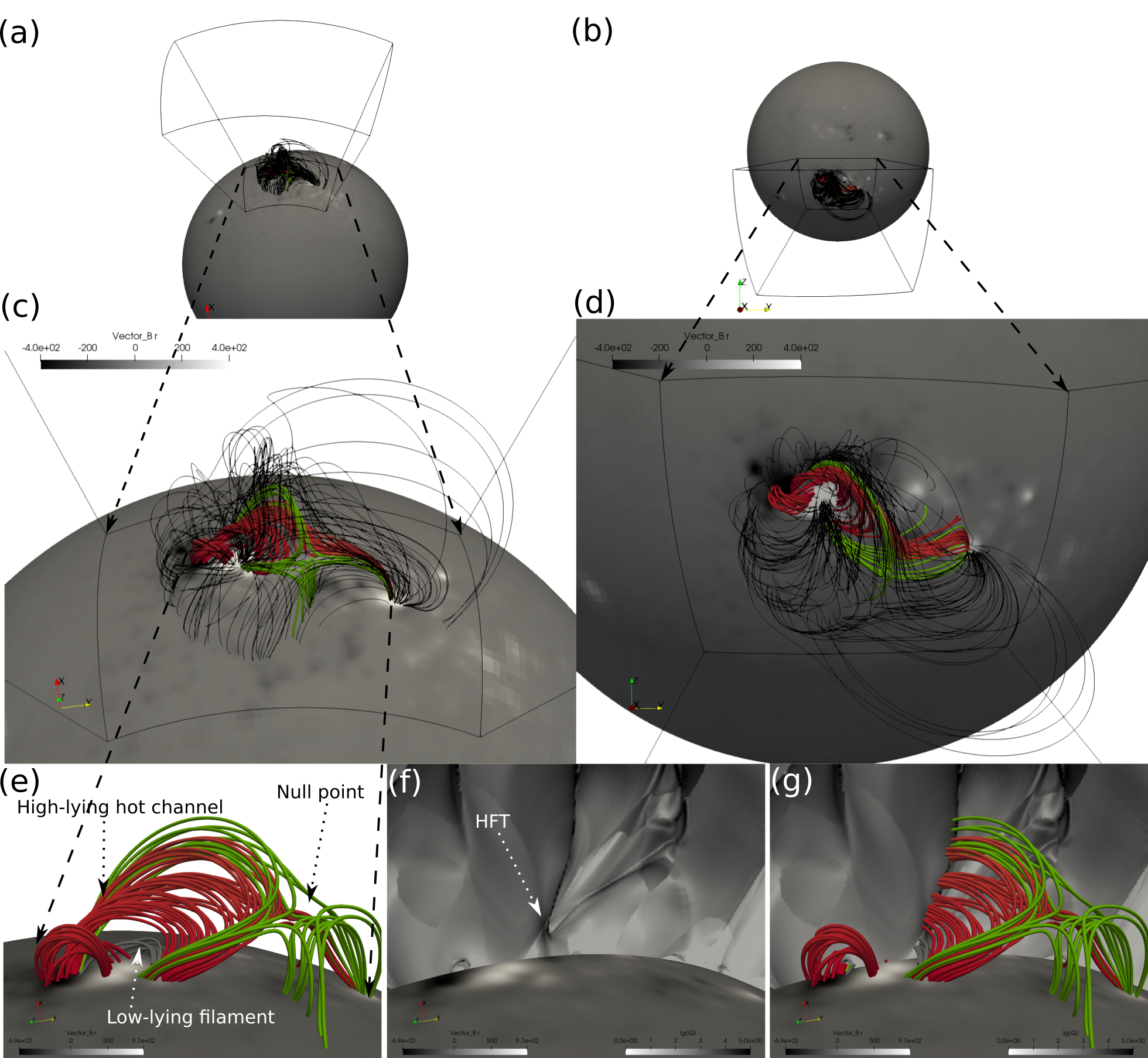}
	\caption{The initial magnetic field configuration in two perspectives. The simulation domain has a grid resolution of ($\approx 0.002 R_{\odot}$) near the bottom boundary, which is colored by radial magnetic field $B_r$. The black field lines in panels (c) and (d) represent the overlying magnetic field rooted at the semi-circular ribbon of high $Q$ factor. The eruptive and stable parts of the MFR are marked in red and gray, respectively, and the green field lines represent the null point structure in panel (e). We show a vertical slice colored by the squashing factor $Q$ and mark the location of the HFT with the white arrow in panel (f). Panel (g) shows that the HFT is located between the magnetic structures of the high-lying hot channel and low-lying filament.}
	\label{fig: f1} 
\end{figure}

The simulation domain is a spherical wedge domain outlined by the black box in panels (a) and (b) of Figure \ref{fig: f1} with $r \in [1.008 R_{\odot}, 2.225 R_{\odot}]$,
$\theta \in [93.4^{\circ}, 125.3^{\circ}]$, and $\phi \in [-15.2^{\circ}, 30.6^{\circ}]$.
The domain is resolved by an $r$-$\theta$-$\phi$ grid of size $396 \times 300 \times 400$ with uniform grid spacing in $\theta$ and $\phi$ and a stretched grid in $r$ where $\Delta r$ varies from 1.5 Mm at the bottom boundary to 3.1 Mm at the top boundary.

\begin{figure}[htb!]
    \centering \includegraphics[width=3in]{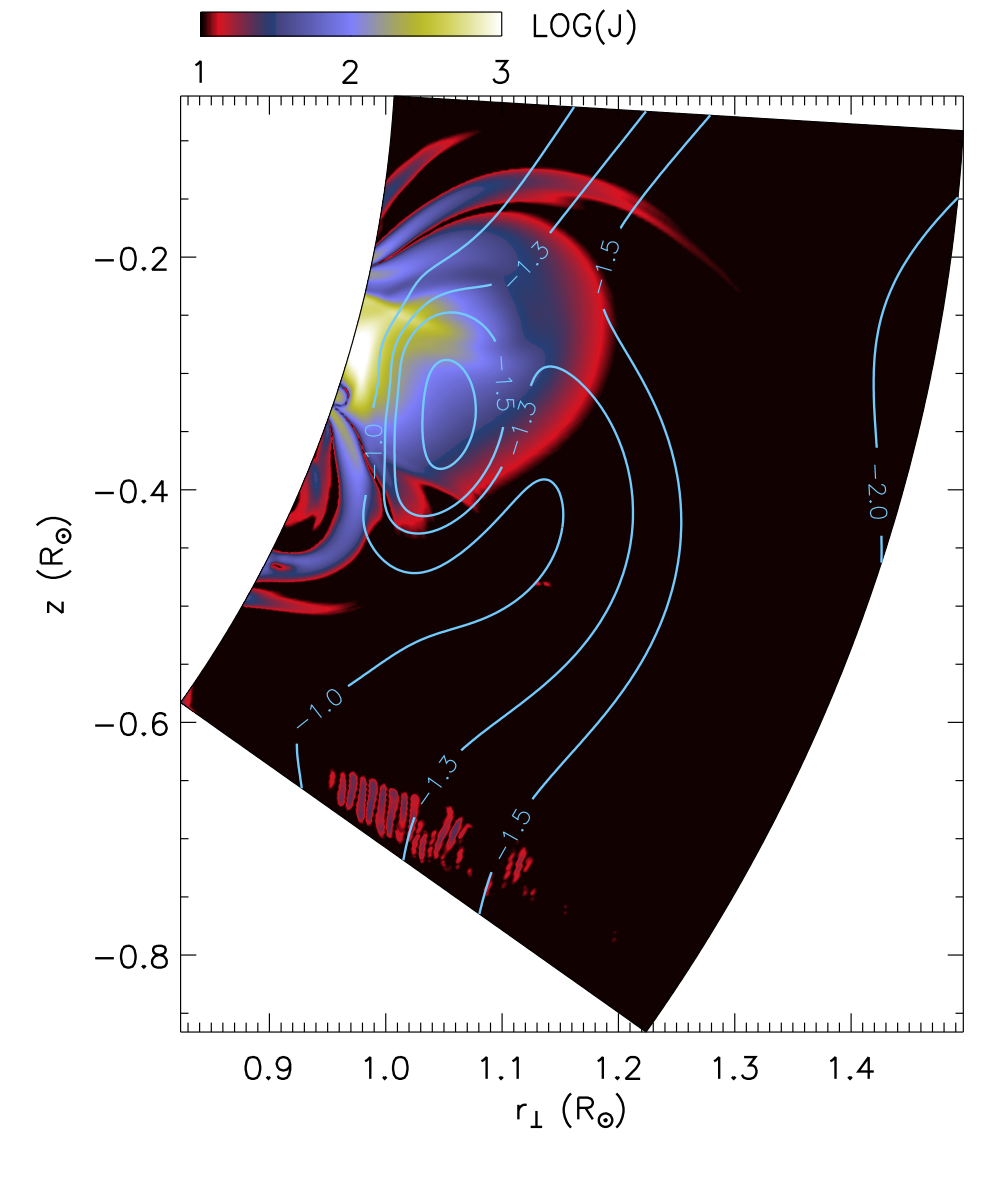}
    \caption{A meridional slice at longitude $\phi=4^{\circ}$ of the current density $J$ overlaid with (cyan) contours of the decay index $d \ln B_p / d \ln h$ of the corresponding potential field for the initial state, where $B_p$ is the potential field strength and $h$ is height above the lower boundary.}
    \label{fig:jc_bpdecay}
\end{figure}

The domain is initialized with a near force-free initial magnetic field solution constructed by the flux rope insertion method for active region 11520 at 2012-07-12 15:00 UT. Detailed descriptions are included in \cite{Liu_2018}. The path of the inserted MFR is the same as that in the first panel of Figure 5 in \cite{Liu_2018}, while the domain size of the current magnetic field model is at least twice as large. The initial axial flux and poloidal flux of the inserted MFR are set to be $4 \times 10^{21} {\rm \ Mx}$ and $6 \times 10^{10} {\rm \ Mx \ cm^{-1}}$. After magneto-frictional relaxation, we obtain the resulting near force-free magnetic field solution, which is an unstable solution, to be used as the initial state of the MHD simulation. \yf{To demonstrate that the initial magnetic field is indeed close to a force-free equilibrium, we have computed the following quantity ${\sigma}_J$, the current weighted average of the sine of the angle between the current and the magnetic field \citep{Wheatland:etal:2000}:
\begin{equation}
   {\sigma}_J = \frac{\sum J_i \sin \theta_i}{\sum J_i}
\end{equation}
where
\begin{equation}
   \sin \theta_i = \frac{|{\bf J} \times {\bf B}|_i}{J_i B_i},
\end{equation}
${\bf J} = \nabla \times {\bf B}$ is the vector current density, $\theta$ is the angle between the $J$ and $B$ vectors, the subscript ``i'' denotes each grid point in the computation domain and the sum is over all the grid points. We obtain $\sigma_J = 0.046 \ll 1$, which means that the angles between the current density vectors and the magnetic field vectors are on average only about ${\sin}^{-1} \sigma_J \approx 3^{\circ}$, i.e. the current density and the magnetic field are nearly parallel, and hence the magnetic field is very close to a force-free equilibrium.
} The initial magnetic field is displayed in panels (c) and (d) of Figure \ref{fig: f1} in which the MFR (red field lines), null point structure (green field lines) and overlying confinement field (black field lines) are shown. In addition to the magnetic field, we also need to initialize the plasma state for the MHD simulation. We assume an initial hydrostatic polytropic atmosphere given by:
\begin{equation}
\rho |_{t=0} = \rho_0 \left [ 1 - \left ( 1- \frac{1}{\gamma} \right )
\frac{GM_{\odot}}{r_0} \frac{\rho_0}{p_0} \left ( 1
- \frac{r_0}{r} \right ) \right ]^{\frac{1}{\gamma - 1}}
\end{equation}
\begin{equation}
p |_{t=0} = p_0 \left [ 1 - \left ( 1- \frac{1}{\gamma} \right )
\frac{GM_{\odot}}{r_0} \frac{\rho_0}{p_0} \left ( 1
- \frac{r_0}{r} \right ) \right ]^{\frac{\gamma}{\gamma - 1}},
\end{equation}
where $r_0$ is $r$ of the lower boundary at the base of the corona, $\rho_0 = 8.365 \times 10^{-16} $ g ${\rm cm}^{-3}$
and $p_0 = 0.219$ dyne ${\rm cm}^{-2}$ are the
initial density and pressure at the lower boundary with the
temperature being $T_0=1.6$ millions of Kelvin (MK).

For the subsequent MHD simulation given the aforementioned initialized magnetic and plasma state, we assume a line-tied lower boundary condition with zero velocity and electric field. The temperature of the first grid zone above the lower boundary is fixed to its initial value of $1.6$ MK, but the density (and hence pressure) is time varying based on the temperature in the grid zone above (being proportional to the temperature in the grid zone above), so as to provide a variable coronal base pressure that increases with the downward heat conduction flux.
For the side and top boundaries, we use simple outward extrapolations that allow the magnetic field and plasma to flow through.

\section{Results} \label{sec: Results}

\begin{figure*}
	\centering \includegraphics[width=7in]{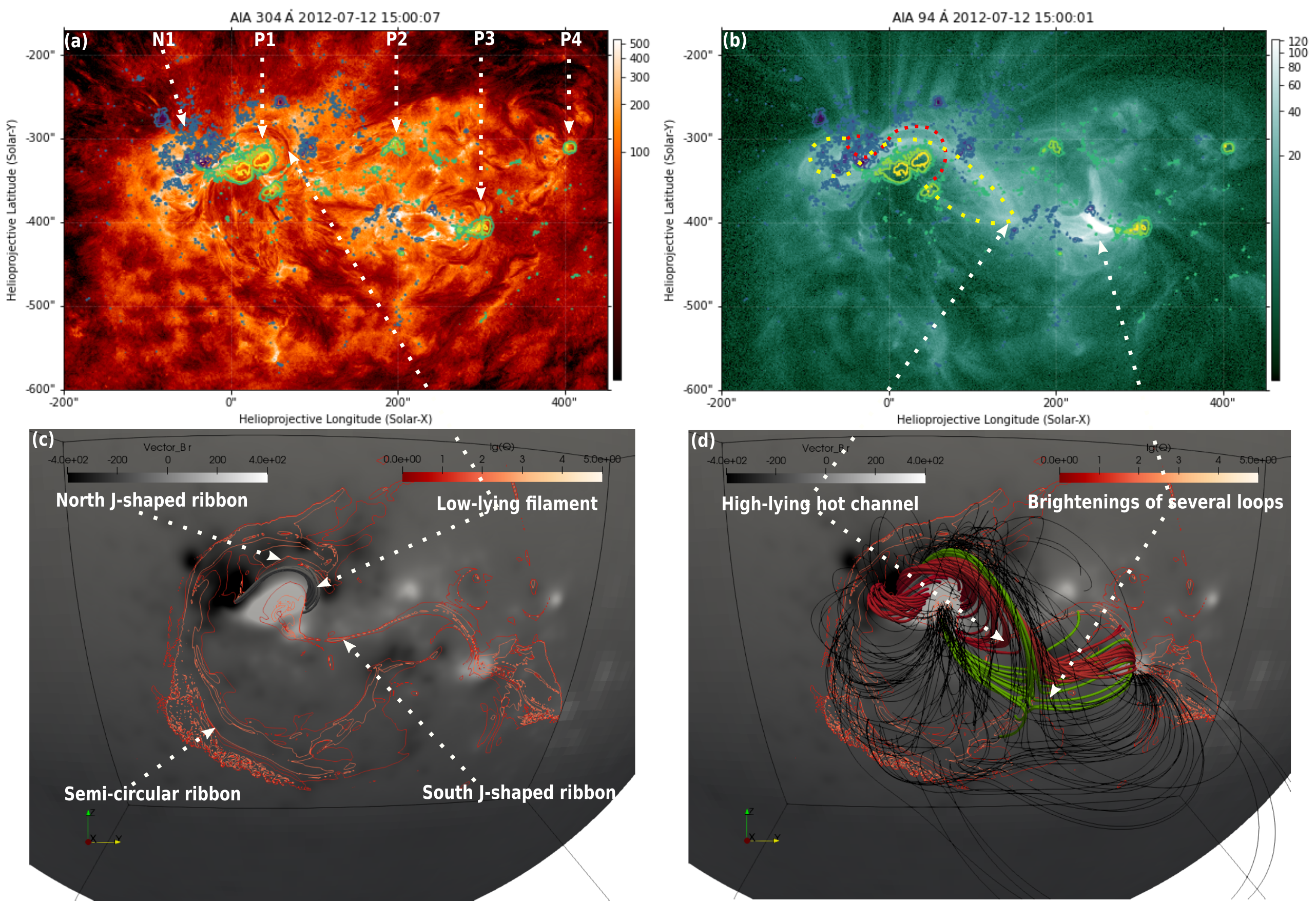}
	\caption{Comparisons between the observations and several characteristic features of the initial magnetic field model. Panel (a) is similar to Figure \ref {fig: f10} (a), but for time $\approx$ 15:00 UT and panel (b) shows the corresponding AIA 94 Å image. In panels (c) and (d), the squashing factor $Q$ and magnetic field lines (same as Figure \ref {fig: f1}) overlaid on $B_r$ are displayed. The observed low-lying filament, high-lying hot channel and brightenings of several loops are compared with the magnetic structures. And the flare ribbons and semi-circular ribbon in Figure \ref {fig: f10} (c) are consistent with the ribbons of high $Q$ factor in panels (c) and (d).} 
	\label{fig: f11} 
\end{figure*}

\begin{figure*}
\begin{interactive}{animation}{f4.mkv}
\centering \includegraphics[width=6.6in]{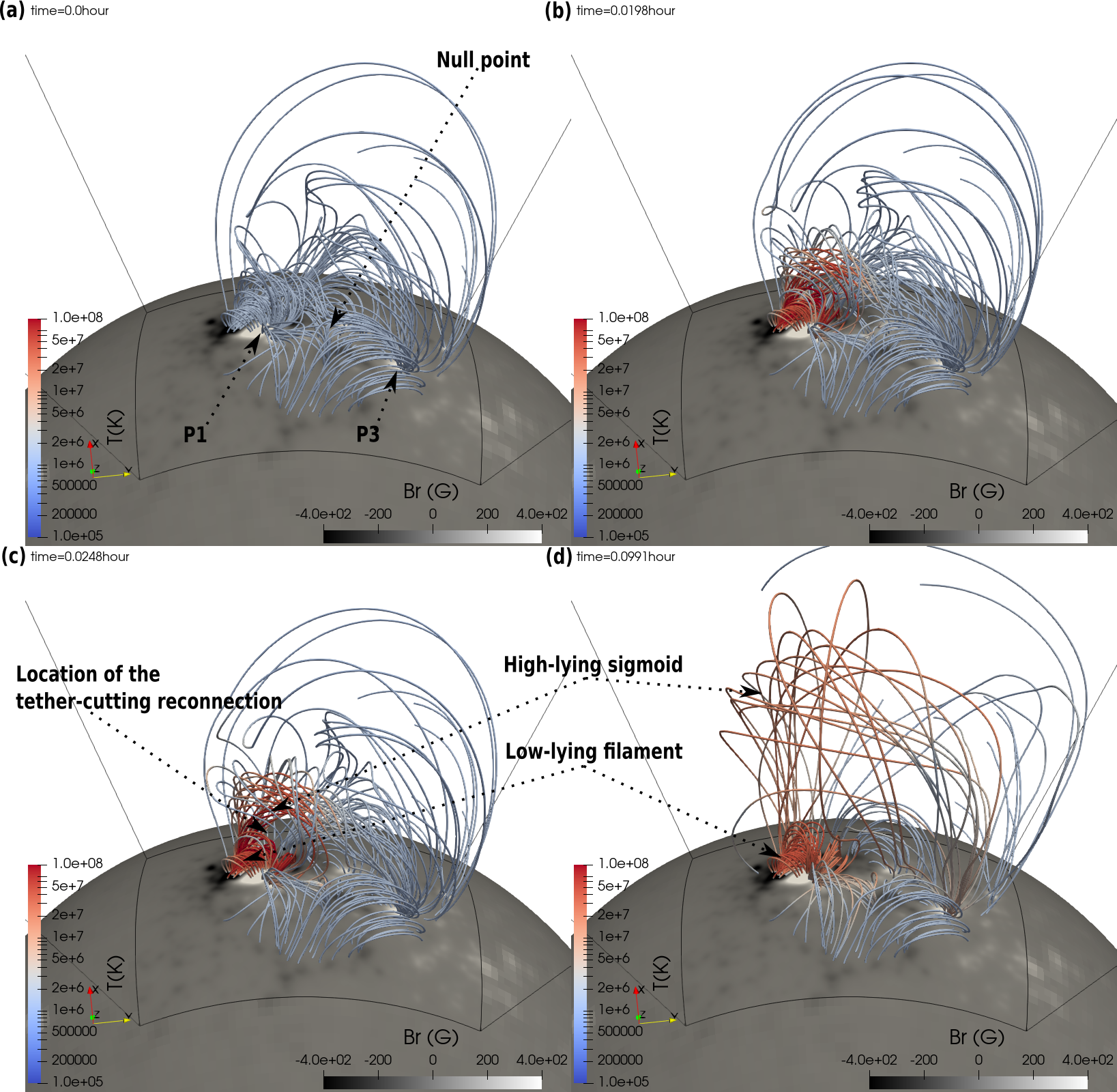}
\end{interactive}
	
	\caption{Eruption of the filament-sigmoid system. Different from Figure \ref {fig: f1}, the magnetic field lines are colored by temperature in Kelvin \ltn{and traced from a set of fixed footpoints.} We highlight the locations of the null point and the two positive magnetic polarities (P1 and P3) with black arrows in panel (a). The high-lying sigmoid, low-lying filament and location of the tether-cutting reconnection are also marked by black arrows in panels (c) and (d). An animated version of this figure is available, which presents the evolution of the filament-sigmoid system from time$=0$ to $0.45$ hour. \ltn{The animation shows the separation of the high-lying hot channel field lines and low-lying filament field lines more clearly.}} 
	\label{fig: f2} 
\end{figure*}

\begin{figure*}
\begin{interactive}{animation}{f5.mkv}
\centering \includegraphics[width=6.6in]{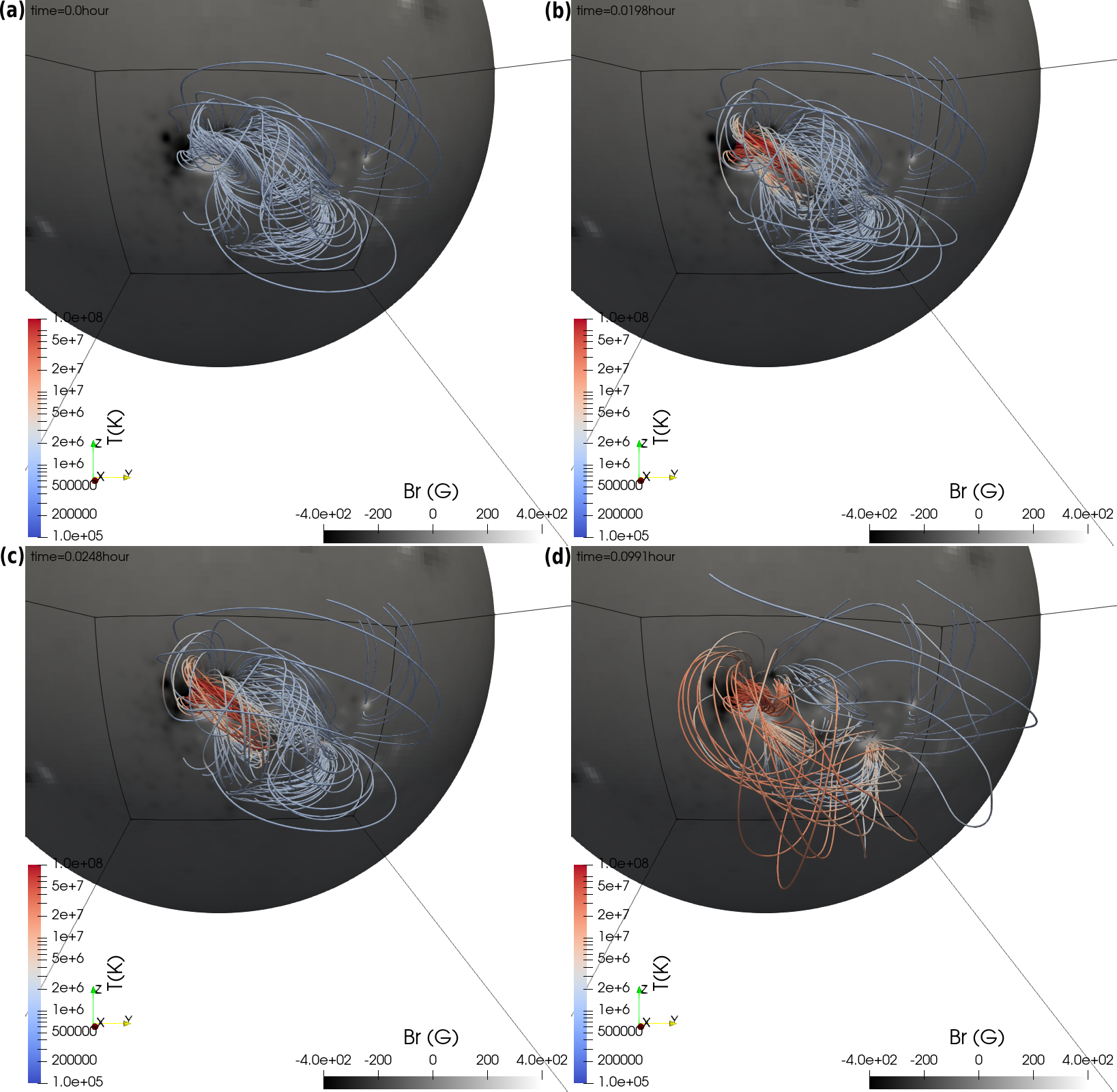}
\end{interactive}
	
	\caption{Same as Figure \ref{fig: f2} but viewed from the Earth position.} 
	\label{fig: f2z} 
\end{figure*}

\begin{figure*}
\begin{interactive}{animation}{f6.mkv}
\centering \includegraphics[width=6.6in]{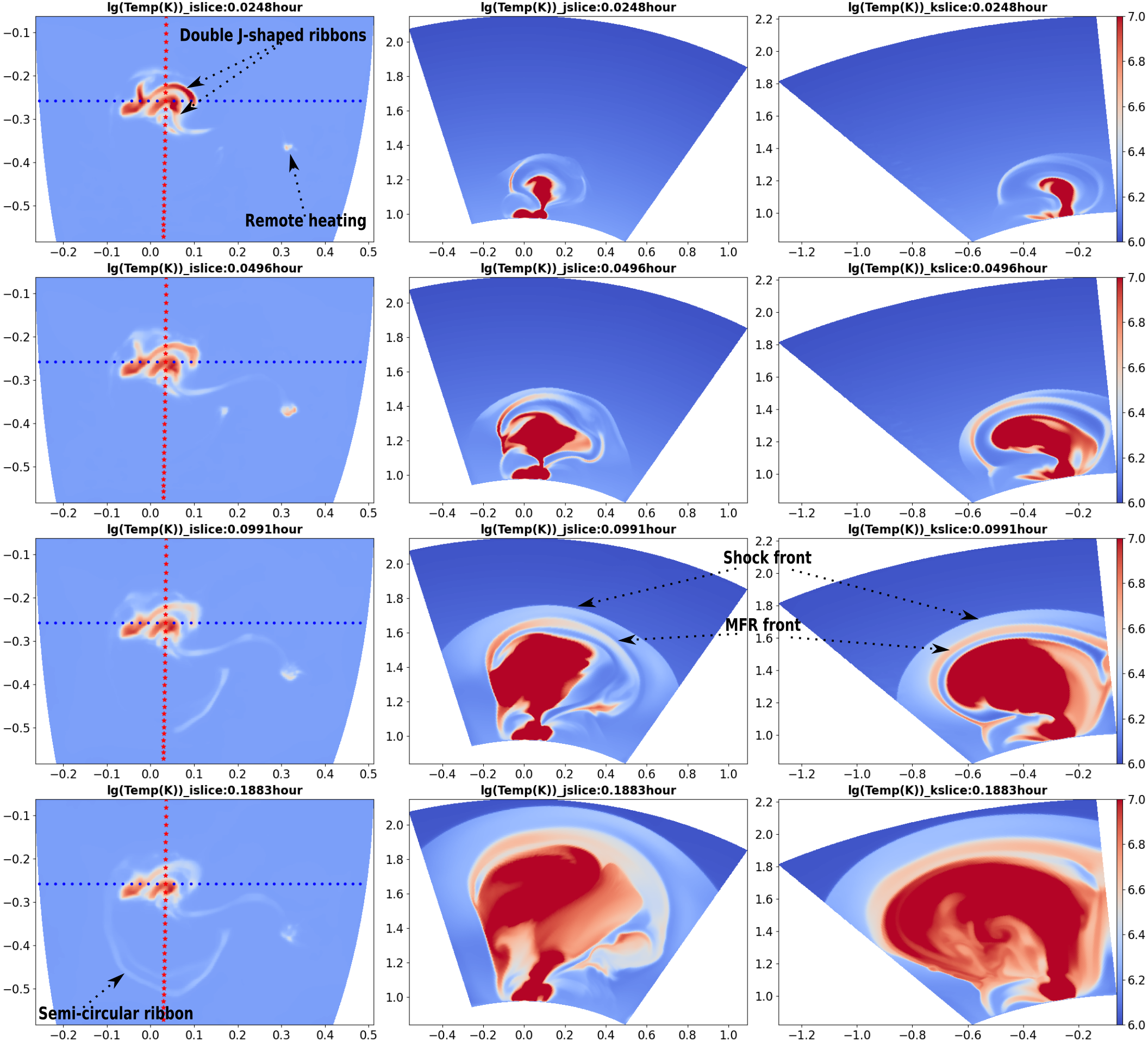}
\end{interactive}
	
	\caption{Temperature distribution on the bottom and two vertical slices. The positions of the vertical slices in the second and third columns are marked by the blue and red dashed lines in the first column, respectively. We mark the double J-shaped ribbons, remote heating, simi-circular ribbon, shock front and MFR front with the black arrows. An animated version of this figure is available, which presents the evolution of the three slices from time$=0$ to $0.45$ hour. \ltn{The evolutions of the marked structures are more clearly visible in the animation.}} 	
	\label{fig: f3}  
\end{figure*}

\begin{figure*}
\begin{interactive}{animation}{f7.mkv}
\centering \includegraphics[width=6.6in]{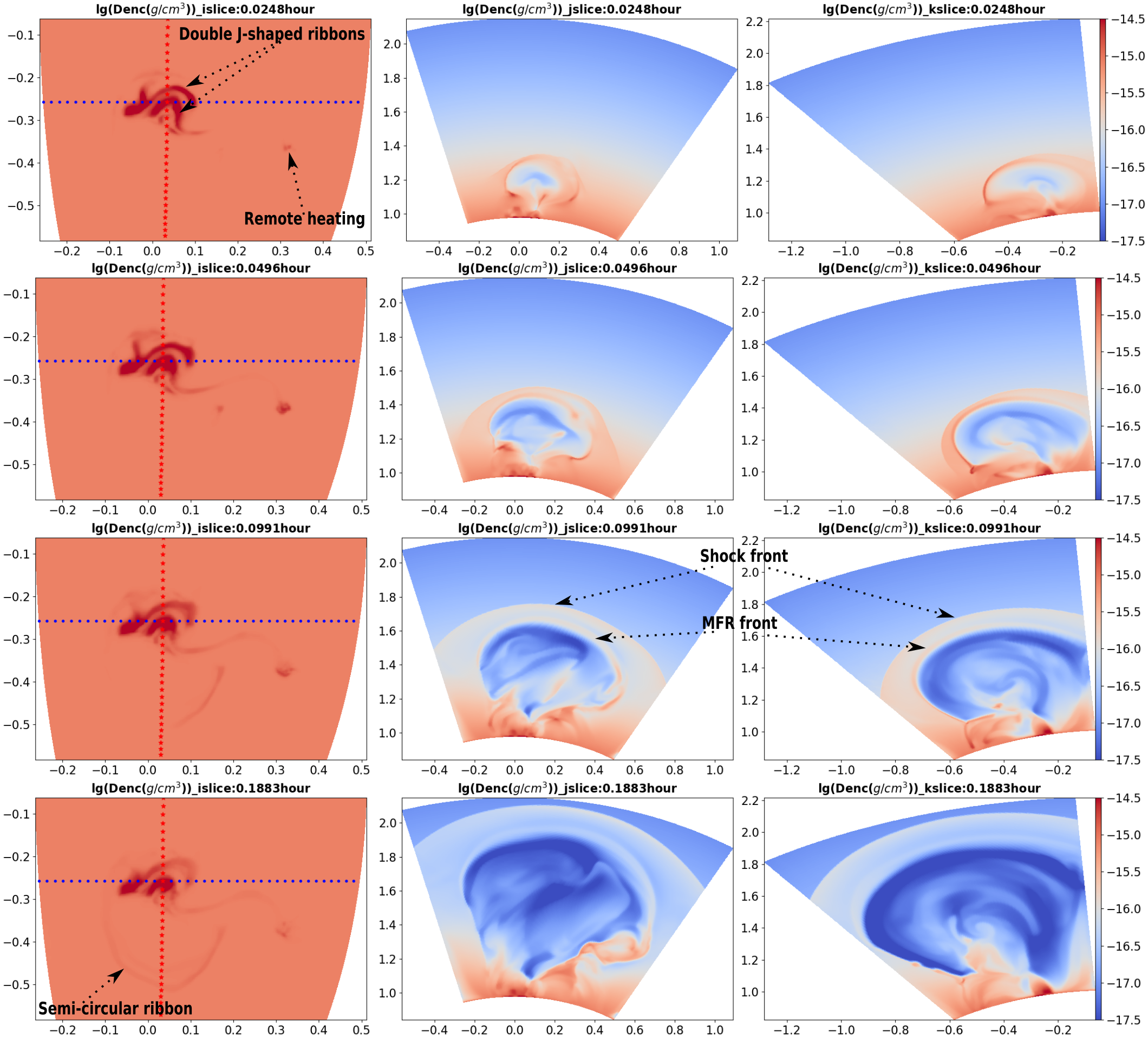}
\end{interactive}
	
	\caption{Same as Figure \ref {fig: f3} but for density.} 	
	\label{fig: f4}  
\end{figure*}

\begin{figure*}
\begin{interactive}{animation}{f8.mkv}
\centering \includegraphics[width=6.6in]{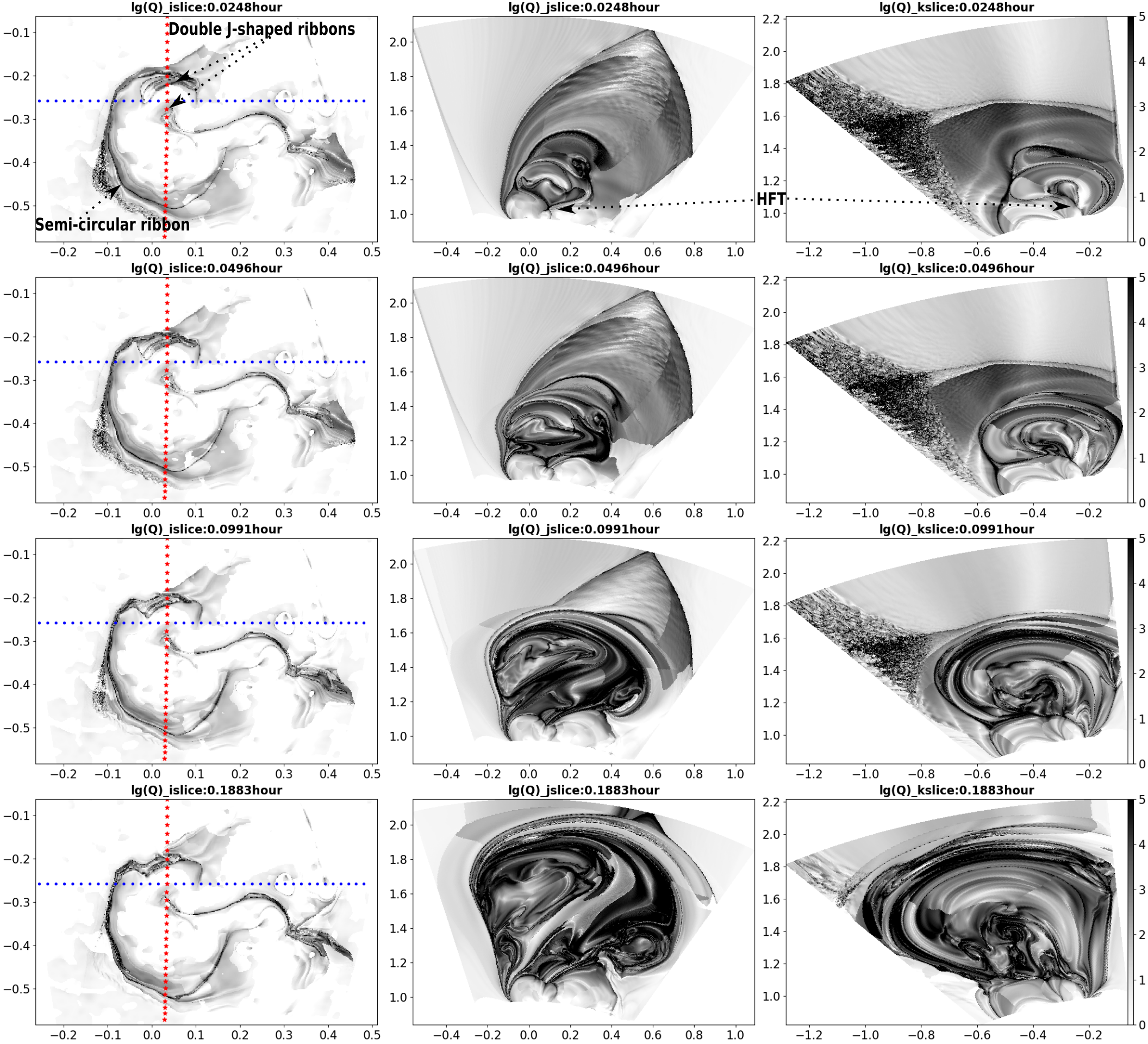}
\end{interactive}
	
	\caption{Same as Figure \ref {fig: f3} but for the squashing factor $Q$. The double J-shaped ribbons, simi-circular ribbon and HFT are marked by the black arrows.} 	
	\label{fig: f9}  
\end{figure*}

\begin{figure*}
\begin{interactive}{animation}{f9.mkv}
\centering \includegraphics[width=6.6in]{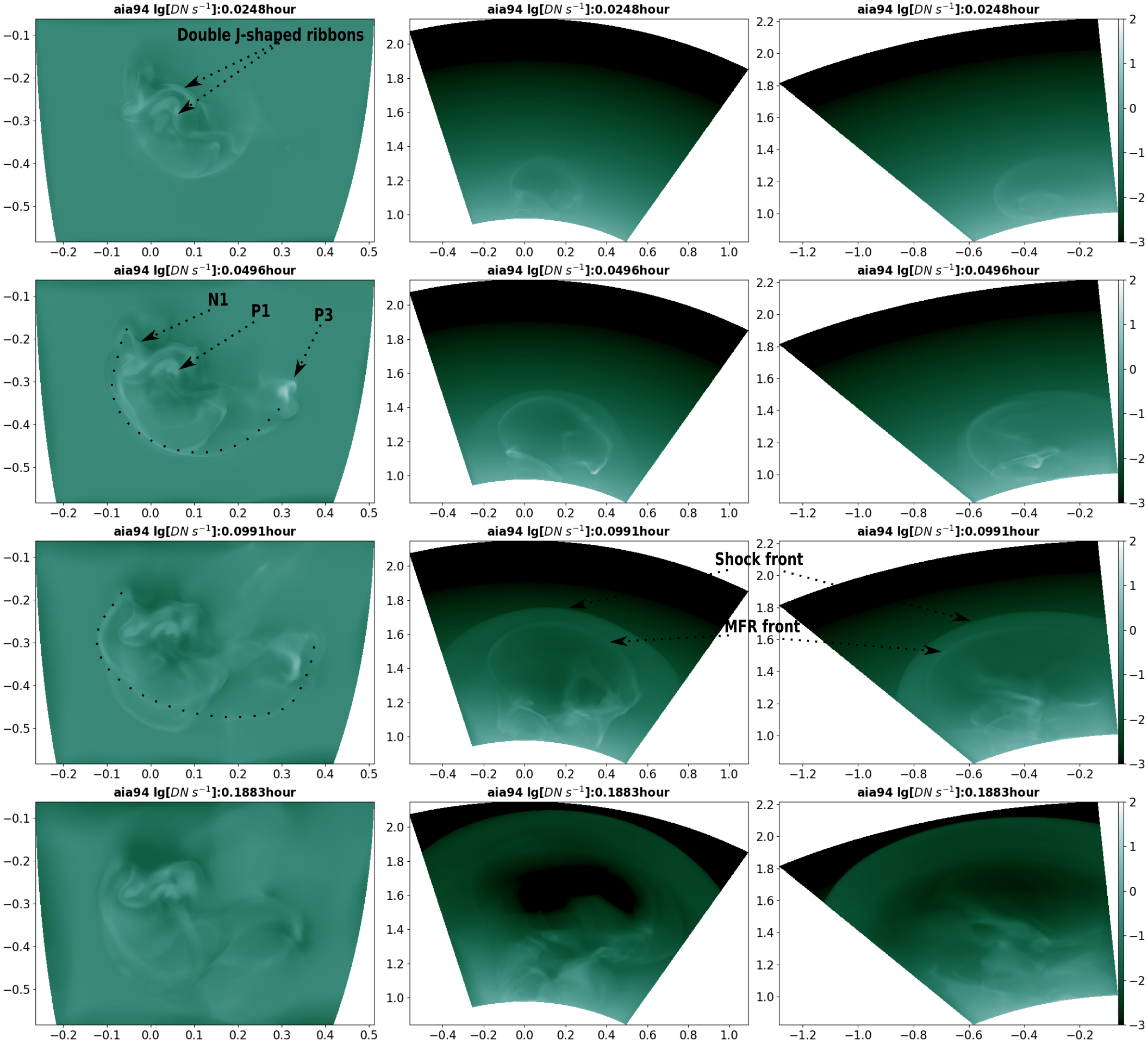}
\end{interactive}
	
	\caption{Synthetic AIA 94 Å images. The images produced by line-of-sight integrations along the x, y and z axes in Figure \ref {fig: f1} are shown in the first, second and third columns, respectively. The double J-shaped ribbons, locations of N1, P1 and P3, shock front and MFR front are marked by the black arrows. We trace the loop-like structure by the black dashed curves. An animated version of this figure is available, which presents the evolution of the synthetic AIA 94 Å images from time$=0$ to $0.45$ hour. \ltn{The evolution of the loop-like structure is easier to identify in the animation.} }
	\label{fig: f5}  
\end{figure*}

\begin{figure*}

\begin{interactive}{animation}{f10.mkv}
\centering \includegraphics[width=6.6in]{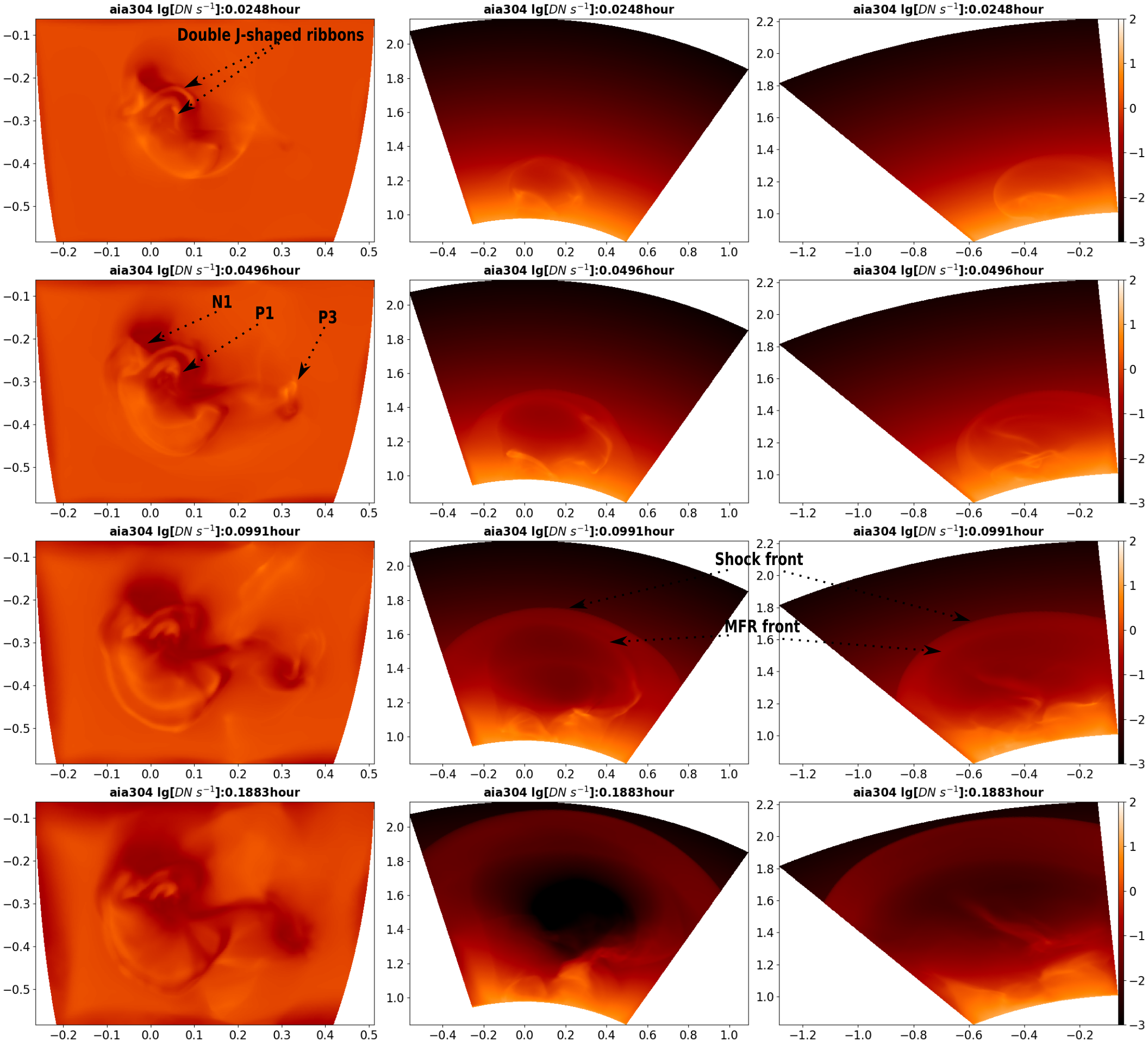}
\end{interactive}
	
	\caption{Same as Figure \ref {fig: f5} but for the synthetic AIA 304 Å images. } 	
	\label{fig: f6}  
\end{figure*}

\begin{figure*}
	\centering \includegraphics[width=7in]{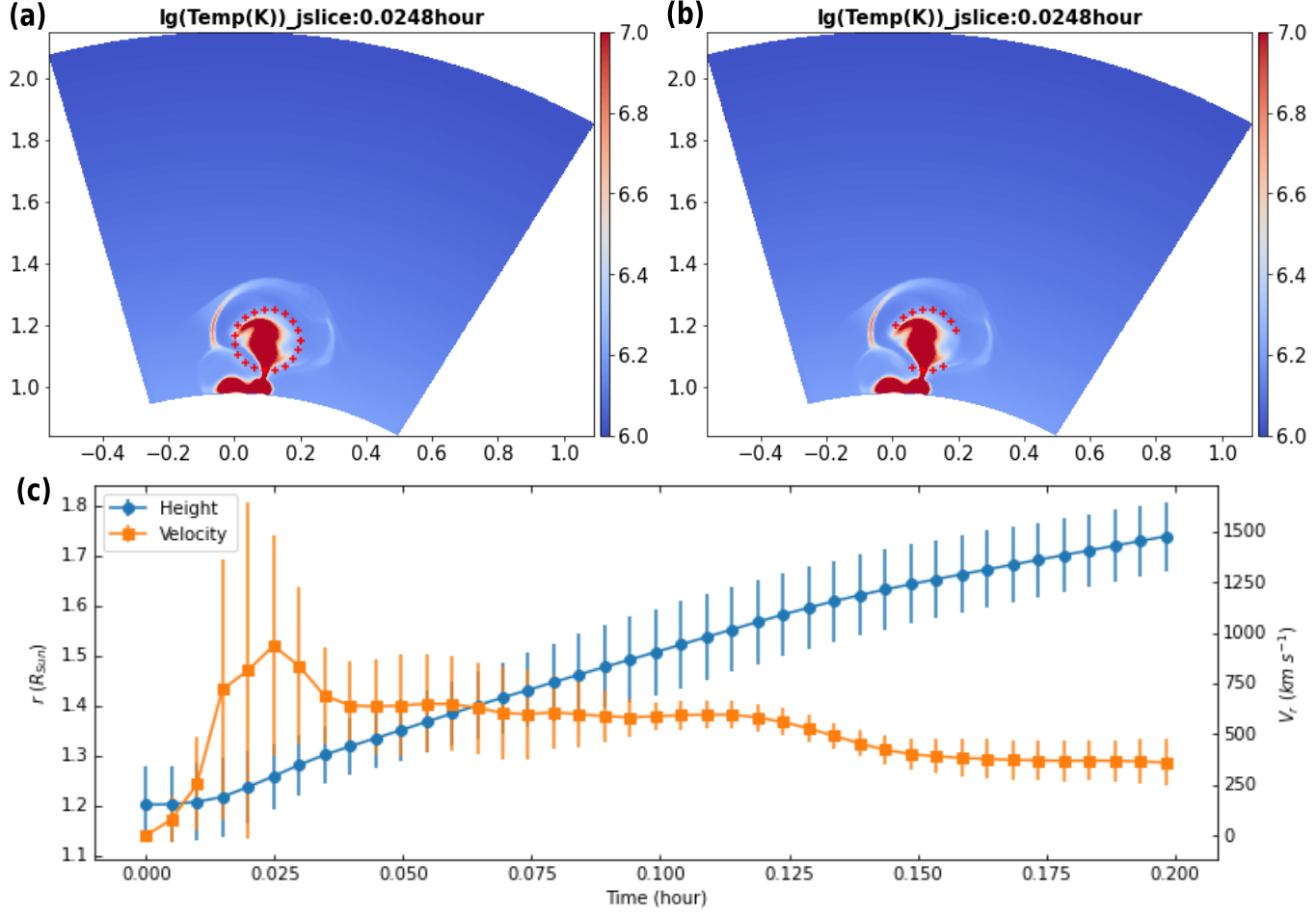}
	\caption{Height and velocity of the erupting MFR with time. We examine the motion of the MFR by tracking 20 mass elements (along the circle) marked by the red `+' in panel (a). 12 of the mass elements in panel (b) erupt with the cavity. We draw the mean heights and velocities of the 12 mass elements, with the standard deviations as the error bars in panel (c).} 	
	\label{fig: f7}  
\end{figure*}

\subsection{Characteristics of the data-constrained MHD simulation} \label{subsec: si}

In this paper, we mainly study the eruption of the filament-sigmoid system through a data-constrained MHD simulation. The initial magnetic model contains a magnetic flux rope (MFR) with an HFT and an overlying null point (panels (c), (d), (e), (f), (g) of Figure \ref {fig: f1}). We find that the magnetic structure can be divided into two parts, above the HFT is the eruption part (red) and below is the stable part (gray) which corresponds to the filament in panels (e), (f) and (g) of Figure \ref {fig: f1}. The null point is located at the right of the MFR and the associated magnetic field lines are colored by green. In addition, the overlying magnetic field lines (black) above the MFR is shown in panels (c) and (d). \yf{Although the initial magnetic field is very close to a force-free equilibrium (as demonstrated in the previous section by evaluating the angle between the current density and the magnetic field), it is an unstable equilibrium. Figure \ref{fig:jc_bpdecay} shows in a meridional slice at longitude $\phi=4^{\circ}$ the current density $J$ overlaid with (cyan) contours of the decay index of the corresponding potential field $d \ln B_p / d \ln h$, where $B_p$ is the potential field strength and $h$ is the height above the lower boundary. It can be seen that a significant portion of the current (or twisted field) is in a region with decay index exceeding -1.5 (see the region enclosed by the -1.5 contour), i.e. subjecting to the onset of the torus instability \citep[e.g.][]{Kliem_2006_PhysicalReviewLetters}. Furthermore, because of the presence of the HFT and the null point, current sheets will readily form driven 
by the unstable rise of the twisted magnetic field. The tether-cutting reconnections in the current sheet that forms at the HFT will further accelerate the twisted field above the HFT. Thus the initial magnetic field is an unstable configuration.} \ltn{Note from Figure \ref{fig:jc_bpdecay} that above the torus-unstable region encircled by the decay index -1.5 contour, there is a torus-stable region where the decay index magnitude goes below 1.5, before it becomes torus-unstable again above the outer contour of decay index -1.5.  Thus the presence of the HFT and the resulting tether-cutting reconnections are important for the continued acceleration of the flux rope across the torus-stable zone \citep{Inoue2018NatCo}.}

Comparisons between the observations and several characteristic features of the initial magnetic model are shown in Figure \ref{fig: f11}. We find that the semi-circular ribbon in Figure \ref{fig: f10}(c) is consistent with the semi-circular ribbon of high squashing factor $Q$ \cite[e.g.][]{Titov:etal:2002} in panels (c) and (d) of Figure \ref{fig: f11}. The flare ribbons correspond to the two J-shaped high $Q$ ribbons around the filament, and the remote brightenings of several loops correspond to the spine-fan structure of the null point. The filament and low-lying sigmoid lie along the edge of the north J-shaped high $Q$ ribbon. Another J-shaped high $Q$ ribbon in the south, extending to P3 is also found, which is consistent with the finding of \cite{Dudik_2014_apj_784_144} that the location of P3 is the rightmost extension of the flare ribbons. The MFR is rooted between the north J-shaped ribbon and the left hook of south J-shaped ribbon, which are consistent with the flare ribbons in Figure \ref {fig: f10}(c). The distribution of the high $Q$ factor in Figure \ref{fig: f11}(c) and magnetic structures in Figure \ref{fig: f11}(d) indicate that there are mainly three magnetic structures, namely, the MFR (red) rooted at the edge of the north J-shaped ribbon and the left hook of the south J-shaped ribbon; the null point structure (green) rooted at the right and left hooks of the south J-shaped ribbon as well as the right hook of the north J-shaped ribbon; and the overlying fan-like structure (black) rooted at the semi-circular ribbon. The corresponding observational characteristics are double J-shaped flare ribbons, brightenings of several loops, and semi-circular ribbon in Figure \ref{fig: f10}.

The simulated eruption of the system is displayed in Figures \ref{fig: f2} and \ref{fig: f2z} as viewed from two different perspectives. We find two kinds of magnetic reconnections, which occur almost simultaneously in the simulation. Considering that the brightenings of several loops around P3 occurs shortly before the eruption of the high-lying sigmoid \citep{Dudik_2014_apj_784_144}, we infer that the overlying reconnection (breakout reconnection proposed by \citealt{Antiochos_1999_apj}) occurs before the tether-cutting reconnection. Based on the results of this simulation, the eruption processes are suggested as follows. First, the overlying reconnection occurs at the null point due to the rise of the unstable MFR, and then the tether-cutting reconnection occurs at the HFT. The overlying reconnection makes the right footpoint of the rising MFR start to move from P1 to P3 in Figures \ref{fig: f2}(a) and \ref{fig: f2}(b), and the tether-cutting reconnection cuts off the connection of the low-lying filament and high-lying sigmoid in Figures \ref{fig: f2}(c) and \ref{fig: f2}(d). In the end, the high-lying sigmoid becomes a loop-like structure in Figures \ref{fig: f2}(d) and \ref{fig: f2z}(d), and erupts as a CME. Strong heating is detected for the erupting MFR and the post-flare loops due to the tether-cutting reconnection at the HFT, as shown by the red field lines in Figures \ref{fig: f2}(b), \ref{fig: f2}(c), and \ref{fig: f2}(d) and Figures \ref{fig: f2z}(b), \ref{fig: f2z}(c), and \ref{fig: f2z}(d). The erupting MFR is heated to about tens of MK at time$ \approx 0.1$ hour in Figures \ref{fig: f2}(d) and \ref{fig: f2z}(d). 
The morphology of the heated erupting MFR in Figure \ref{fig: f2z}(b) is consistent with that of the high-lying sigmoid in Figure \ref{fig: f10}(f), and the heated post-flare loops below the erupting flux rope in Figure \ref{fig: f2z}(d) are consistent with the post-flare loop brightening shown in Figures \ref{fig: f10}(g) and \ref{fig: f10}(h).

We show the evolution of temperature and density on three slices of the simulation domain in Figures \ref{fig: f3} and \ref{fig: f4}. The left column panels show the evolution on the bottom slice. We see the double J-shaped ribbons of enhanced temperature and density on the bottom slice, which correspond to the footpoints of the heated field lines undergoing tether-cutting reconnection, and are proxies of the flare ribbons. Their morphology agrees qualitatively with that of the flare ribbons shown in Figures \ref{fig: f10}(c) and \ref{fig: f10}(d). We also find a transient remote heating site that becomes evident from $time \approx 0.025$ hour to $time \approx 0.05$ hour and weakens later gradually. This remote heating site is activated by the overlying reconnection at the null-point. The semi-circular ribbon is detected and enhances from $time \approx 0.10$ hour to $time \approx 0.19$ hour, which is a signature of magnetic reconnection between the erupting MFR and overlying confinement fields, as expected by the presence of the high Q semi-circular ribbon at the footpoints of the overlying confinement field shown in Figures \ref{fig: f11}(c) and \ref{fig: f11}(d) for the initial state. The locations of the semi-circular ribbon and remote heating shown in the left columns of Figures \ref{fig: f3} and \ref{fig: f4} are consistent with that of the semi-circular ribbon and brightenings of several loops detected in Figures \ref{fig: f10}(c) and \ref{fig: f10}(f). From the two vertical slices in the middle and right columns in Figures \ref{fig: f3} and \ref{fig: f4}, we find a cavity with high temperature and low density that corresponds to the erupting MFR. The distributions of the temperature and squashing factor $Q$ in Figures \ref{fig: f3} and \ref{fig: f9} suggest that the heating is caused by the overlying reconnection, tether-cutting reconnection and magnetic reconnection within the flux rope (internal reconnection) similar to \cite{Guo2012ApJ} and \cite{Zhong2021NatCo}. With the eruption of the MFR, a fast shock front with a dense sheath develops in front of the cavity (MFR) as can be seen in the middle and right columns of Figures \ref{fig: f3} and \ref{fig: f4}.

Figure \ref{fig: f9} shows the squashing factor $Q$ in the corresponding three slices as those in Figures \ref{fig: f3} and \ref{fig: f4}. We find that the double J-shaped structures in Figures \ref{fig: f3} and \ref{fig: f4} are consistent with the high $Q$ J-shaped ribbons, which are  further surrounded by a semi-circular high $Q$ ribbon. The location of the remote heating corresponds to the high $Q$ region of the null point structure. During the eruption, the semi-circular high $Q$ ribbon keeps stationary, while the double J-shaped high $Q$ ribbons separate from each other, which is clearer in the corresponding animated version of Figure \ref{fig: f9}. We find that there are corresponding signatures of heating along the semi-circular high $Q$ ribbon in the first column of Figure \ref{fig: f3}, which suggests the occurrence of magnetic reconnection between the erupting MFR and the overlying magnetic field. The second and third columns of Figure \ref{fig: f9} show the Q maps on the cross-sections of the erupting MFR. We note that the magnetic structure below the HFT is quite stable without obvious morphology change, which is consistent with the stable sheared arcades and post-flare loops in Figure \ref{fig: f2}.

\subsection{Synthetic AIA images} \label{subsec: aia}
\lt{We calculate synthetic AIA 94 and 304 Å images from optically thin coronal plasma on 3D grids, similar to \cite{Doorsselaere2016}. The AIA emission is a function of the density, temperature and AIA temperature
response. We obtain the final synthetic AIA image by integrating the emission along the line of sight.}
The synthetic AIA 94 Å and 304 Å images are displayed in Figures \ref{fig: f5} and \ref{fig: f6}. Note, we cannot model the synthetic EUV emission for the pre-eruption field because the initial plasma state specified for the MHD simulation is a simple 1D hydro-static atmosphere and the initial magnetic field is alread unstable.  We here only model the synthetic EUV emission features produced by the heating resulting from the eruption of the magnetic field. Similar to Figure \ref{fig: f3}, the bright double J-shaped ribbons and brightenings of several loops are detected in the first column. In addition, we detect a loop-like structure traced by the dashed black curves in AIA 94 Å at $time \approx 0.050$ and $0.099$ hours during which the right footpoint of the structure moves westward to P3, which is consistent with the erupting MFR in Figure \ref{fig: f2} and corresponds to the high-lying hot channel in Figure \ref{fig: f10}. We do not detect the corresponding loop-like structure in the synthetic AIA 304 Å. This observation characteristic is similar to the hot channel, which can be observed in high temperature passbands but is absent in low temperature passbands, for example, the observations of panels (b) and (f) of Figure \ref{fig: f10} and observations in \cite{Cheng_2016_apjs_225_16}. It suggests that the traced loop-like structure corresponds to the erupted hot channel or sigmoid. At the same time, a bright front and dark cavity are displayed in the second and third columns. We find that the bright front is the MFR front in the second and third columns of Figure \ref{fig: f5}, and is more obvious in the synthetic AIA 94 Å than in the corresponding synthetic 304 Å images in the second and third columns in Figure \ref{fig: f6}. The bright front in the side view (second and third columns of Figure \ref{fig: f5}) corresponds to the loop-like structure in the overhead view (first column of Figure \ref{fig: f5}). The data-constrained simulation confirms that the observed hot channel is the proxy of the MFR, proposed by observational analyses such as \cite{Zhang_2012_NatureCommunications_3_747} and \cite{Cheng_2013_apj_763_43}. Furthermore, the simulation indicates that the loop-like structure seen later after the eruption of the hot channel is consistent with the MFR front.

Compared with the classical three-component CME, the bright core is absent in the simulation, which is due to the absence of the filament eruption and is consistent with the observations of \citet{Cheng_2014_apj_789_93}. In this simulation, we construct an unstable magnetic configuration as the initial magnetic field. We do not include the optically thin radiative cooling that can drive the formation of the cool prominence condensations, so the simulation precludes the formation of the dense prominence. Thus our simulation can not directly address the observed absence of an erupting filament in the CME. It only shows the separation of the initial MFR and sheared arcades due to the magnetic reconnection at the HFT, and if the filament plasma is confined below the HFT, it would explain the absence of an erupting filament.

In Figure \ref{fig: f7}, we show the time-height and time-velocity plots of the eruption. We choose 20 mass elements along a circle in the cavity in panel (a), and trace the positions and speeds of the 20 mass elements. 8 mass elements at the two sides are squeezed to the two flanks of the MFR and do not erupt. The remaining 12 mass elements in panel (b) erupt with the MFR. The mean heights and velocities of the 12 mass elements are regarded as the heights and velocities of the erupted MFR, which are drawn in panel (c) with the standard deviations as the error bars. We find that the velocity increases rapidly in the beginning, then decreases slowly, and then reaches a uniform speed of about 370 $km/s$ when the MFR reaches the height of about 1.7 $R_{Sun}$ at $time \approx 0.2$ hour. Observations such as \cite{Zhang2001ApJ} and \cite{Cheng2020ApJ} indicate that the eruption of the MFR typically include three phases: slow rise, fast rise (impulsive acceleration) and propagation. The slow rise phase is not seen in the simulation because the initial magnetic field is already unstable at the beginning.


\section{Summary and Discussion} \label{sec: Summary}

In this paper, we investigate the magnetic field evolution that results in the separation between the low-lying filament and high-lying sigmoid in a data-constrained MHD simulation of the flare/CME event in solar active region 11520 on July 12, 2012. Observations in AIA 304 Å and 94 Å show that the filament-sigmoid system contains a low-lying filament and a high-lying sigmoid. An additional low-lying sigmoid is detected in this event. \cite{Cheng_2014_apj_789_93} suggest that the magnetic structure of the two sigmoids is a double-decker MFR system where the two sigmoids correspond to two MFRs overlying on the same PIL. While \cite{Liu_2018} propose that the magnetic structure of this filament-sigmoid system is consistent with an HFT-MFR system, where the high-lying sigmoid corresponds to the MFR above the HFT, the low-lying filament is located in the dips of the sheared arcades below the HFT. And the observed low-lying sigmoid is not a hot channel, it corresponds to the flare ribbon heated by energetic particles and plasma produced by the tether-cutting reconnection at the HFT. However, we can not rule out the possibility of the magnetic structure of this filament-sigmoid system being double-decker MFRs, where the separation between the low-lying MFR and high-lying MFR can also be reproduced \citep{Kliem2014ApJ}. In addition, the separation looks similar to the partial eruption \citep{Gilbert2000ApJ,Gilbert2001ApJ,Gibson2006ApJ,Gibson2008JGRA} where the sigmoid and filaments are located in the MFR and coupled with each other \citep{Gibson2006JGRA}. Magnetic reconnections at current sheets make the MFR break into two during the partial eruption. However, \lt{the filament-sigmoid system studied in this work possesses separated sigmoid and filaments before eruption.}  Our findings and conclusions apply to this situation.

The eruption mechanism is difficult to determine in this simulation, because the initial magnetic field is an unstable model. In the simulation, the overlying reconnection occurs almost simultaneously with the tether-cutting reconnection. We find that the overlying reconnection at the null point drives the right footpoint of the high-lying sigmoid to move from P1 to P3 during which the remote heating is detected. The tether-cutting reconnection at the HFT leads to the separation of the low-lying filament and high-lying sigmoid. The high-lying sigmoid erupts, forming a CME without clear bright core, due to the fact that the magnetic structure of the low-lying filament stays stable, which is consistent with the observations of \cite{Cheng_2014_apj_789_93}. During the eruption processes, we compare the observations with simulation and find that the brightenings of several loops in AIA 94 Å around P3 are related to the overlying reconnection at the null point, the flare ribbons correspond to the double J-shaped high $Q$ ribbons and the semi-circular ribbon is along the semi-circular high $Q$ ribbon. 
\lt{The rising of the unstable magnetic flux rope squeezes the null point, triggers the overlying reconnection and the tether-cutting reconnection at the HFT. The magnetic structure and evolution of the filament-sigmoid system indicate that the post flare loops cover the low-lying filament and make it more stable when the tether-cutting reconnection occurrs at the HFT. This provides a very intuitive explanation for why the filament can survive violent solar eruptions in some cases.}



We calculate synthetic AIA 94 Å and 304 Å images and find that a loop-like structure in the synthetic AIA 94 Å images corresponds to the front of the erupting MFR, whose right footpoint moves form P1 to P3 during the eruption. And in the side views, we find that the bright loop-like structure matches the upper boundary (front) of the MFR with high density and temperature. We do not find the corresponding signature of the loop-like structure in the synthetic AIA 304 Å images, so the loop-like structure is likely consistent with the observed hot channel, which is visible in AIA 94 Å but not AIA 304 Å.
The counterpart in the synthetic EUV emission of the loop-like structure before the eruption is not modeled due to the unstable initial magnetic model in the simulation. The loop-like structure corresponds to the hot channel in the rise phase and is the upper front of the erupted MFR. The whole erupted MFR composed of the dark cavity and brignt front (loop-like structure) in the synthetic images, is heated by the overlying reconnection, tether-cutting reconnection and internal reconnections in the simulation. The MFR front appears as a bright loop-like structure, while the cavity is dark in the synthetic AIA 94 Å images, due to the low density of the cavity even though it is heated and is of high temperature.


No prominence condensation is presented in this simulation, because the simulation does not include the optically thin radiative cooling. Furthermore, the initial magnetic field is unstable and quickly erupts with strong heating produced by the reconnection at the HFT, so that even if the optically thin radiative cooling was included, the prominence condensations might not be able to form. The simulation reproduces the separation of an erupting flux rope (corresponding to the erupted hot channel) and a stable lower sheared arcade (corresponding to the observed stable lower filament), which is consistent with the observations \citep{Cheng_2014_apj_789_93}. In summary, the observational characteristics such as the erupted hot channel, brightenings of several loops, flare ribbons, and semi-circular ribbon are reproduced in this data-constrained MHD simulation. 
\lt{The separation of the flare ribbons is a common phenomenon while in this case the separation is not very prominent in the observation and simulation, which is shown in Figures \ref{fig: f3}--\ref{fig: f6}, especially the corresponding animations.}

The simulation morphologically reproduces observational characteristics, but the height-time and velocity-time curves of the erupting MFR only include the impulsive acceleration and propagation phases. The initial magnetic model is already unstable, which leads to the absence of the slow rise phase in the simulation. To test this conjecture, simulations with a stable initial magnetic models are needed, which is our future plan. This data-constrained simulation gives a comprehensive understanding of the complex magnetic field evolution that can lead to the observed eruption of the filament-sigmoid system in AR 11520.

We thank the anonymous referee for helpful comments that improved the paper. We thank Sarah
Gibson for reading the paper and helpful comments. We sincerely acknowledge Dr. Adriaan van Ballegooijen who not only developed the flux rope insertion method but also provided us with valuable suggestions when combing the magnetic model with the magnetic flux eruption code. SDO is a mission of Living With a Star Program from NASA.
Tie Liu is supported by the National Natural Science Foundation of China grant NO. 11773016, 11473071, 11790302 (11790300), 41761134088, and U1731241. Y. Fan's work is supported by the National Center for Atmospheric Research (NCAR), which is a major facility sponsored by the National Science Foundation under Cooperative Agreement No. 1852977. Y. Fan also acknowledges high-performance computing support from Cheyenne (doi:10.5065/D6RX99HX) provided by NCAR's Computational and Information Systems
Laboratory, sponsored by the National Science Foundation. Y.G. is supported by NSFC (11773016, 11733003, and 11961131002) and 2020YFC2201201. Yingna Su is supported by the National Key R\&D Program of China 2021YFA1600502 (2021YFA1600500) , the Chinese foundations NSFC (12173092, 41761134088,
11790302 (11790300)), and the Strategic Priority Research Program on Space Science, CAS, Grant No. XDA15052200 and XDA15320301.


\bibliographystyle{aasjournal}
\bibliography{refs2022}
\end{document}